 \documentclass[aps,pra,twocolumn,superscriptaddress]{revtex4}

\usepackage{amsmath,amssymb,amsthm,color,mathrsfs}
\usepackage{amsbsy,mathtools}

\usepackage{hyperref}
\usepackage{graphicx}

\newcommand{\ox}{\omega_X}
\renewcommand{\oc}{\omega_C}
\newcommand{\wx}{\varpi_X}
\newcommand{\wc}{\varpi_C}
\newcommand{\kx}{\kappa_X}
\newcommand{\kc}{\kappa_C}

\newcommand{\oLP}{\omega_{\rm LP}}

\newcommand{\exciton}{{\psi_X}}
\newcommand{\photon}{{\psi_C}}
\newcommand{\excitonenv}{\phi_X}
\newcommand{\photonenv}{\phi_C}

\begin{document}

\title{Continuous and discontinuous dark solitons in polariton condensates}
\author{Stavros Komineas}
\affiliation{Department Applied Mathematics, University of Crete, 71409 Heraklion, Crete, Greece}
\author{Stephen P. Shipman}
\affiliation{Department of Mathematics, Louisiana State University, Baton Rouge, Louisiana \ 70803, USA}
\author{Stephanos Venakides}
\affiliation{Department of Mathematics, Duke University, Durham, North Carolina \ 27708, USA}
\begin{abstract}
Bose-Einstein condensates of exciton-polaritons are described by a Schr\"odinger system of two equations.
Nonlinearity due to exciton interactions gives rise to a frequency band of dark soliton solutions, which are found analytically for the lossless zero-velocity case.
The soliton's far-field value varies from zero to infinity as the operating frequency varies across the band.
For positive detuning (photon frequency higher than exciton frequency), 
the exciton wavefunction becomes discontinuous when the operating frequency exceeds the exciton frequency.
This phenomenon lies outside the parameter regime of validity of the Gross-Pitaevskii (GP) model.
Within its regime of validity, we give a derivation of a single-mode GP model from the initial Schr\"odinger system and compare the continuous polariton solitons and GP solitons using the healing length notion.
\end{abstract}

\pacs{71.36.+c,42.65.Tg,05.45.Yv}

\maketitle

\section{Introduction}

Exciton-polaritons are matter-light quasiparticles that arise from the coupling between excitons and photon modes in a semiconductor microcavity and can form Bose-Einstein condensates (BEC) at relatively high temperatures
\cite{KavokinBaumberg2007,CarusottoCiuti_RMP2013,DengHaug_RMP2010,KeelingMarchetti_SST2007}.
Polariton condensates are sustained by laser pumping of photons in a two-dimensional quantum well.
In a mean-field approximation,  their wavefunctions produce a rich variety of localised quantum states
in the micrometer scale: dark solitons \cite{LarionavoaStolz_OptLett2008,AmoPigeon_Science2011,PigeonCarusottoCiuti_PRB2011,HivetFlayac_NatPhys2012,GrossoNardin_PRB2012}, bright solitons \cite{LarionavoaStolz_OptLett2008,SichKriszanovskii_NatPhot2012,EgorovSkryabin_PRL2009,EgorovGorbach_PRL2010}, vortices \cite{MarchettiSzymanska,GrossoNardin_PRL2011}.
Solitons in polaritonic condensates have potential for applications in ultrafast information processing
\cite{AckemannFirthOppo_2009} due to picosecond response times and strong nonlinearities
\cite{SichKriszanovskii_NatPhot2012,EgorovSkryabin_PRL2009}.

In this work, we report a frequency band of dark polariton solitons whose exciton wave function develops a discontinuity as the frequency is increased beyond the exciton frequency (Fig.~\ref{fig:solitons}).  At the point of discontinuity, the photon field vanishes while the exciton field experiences a half-cycle phase jump.

We investigate a one-dimensional condensate of polaritons in a strongly-coupled exciton and photon system.  Our derivation depends crucially on the use of the classic model that retains separate wave functions for the excitons and the photon modes.  Exciton interactions are modelled by a nonlinear term, while photons are dispersive.  Neglecting both pumping and losses (which are due to radiation and thermalization) and thus focusing on the synergy of exciton interaction (nonlinearity) and photon dispersion allows us to produce analytical formulae
for polariton solitons.
Conservative solitonic structures of half-light and half-matter have been considered in the literature
\cite{SpatialSolitons_2001}.

The  solitons we derive apply for a short time after the pumping is removed and the losses have not seriously manifested themselves or the solitons lie outside the pump spot.
For example, in Refs.~\cite{AmoPigeon_Science2011,GrossoNardin_PRB2012}
quasi-one-dimensional structures are observed outside the pump spots.
In a different realization, polariton condensates can be created at two pump spots \cite{TosiChristmannSavvidis_NatPhys2012}
and localised structures can be sustained in the region between the two spots where there is no pumping.

\section{Polariton Solitons}

We consider a one-dimensional semiconductor microcavity in which a photon field $\photon(x,t)$ interacts with an exciton field $\exciton(x,t)$. One dimensional or nearly one dimensional polariton structures have been observed in
\cite{AmoPigeon_Science2011,GrossoNardin_PRB2012,TosiChristmannSavvidis_NatPhys2012} and \cite{DreismannCristofolini_2014} (radial fields). The pair $(\exciton,\photon)$ is a polariton field and is modeled by the system
\cite{KavokinBaumberg2007,Deveaud2007,CarusottoCiuti_PRL2004,YulinEgorov_PRA2008}
\begin{subequations}\label{eq:polariton}
\begin{align}
  i\partial_t\exciton &= \left( \ox-i\kx + g|\exciton|^2 \right) \exciton + \gamma\photon \label{eq:polaritonA}\\
  i\partial_t\photon &= \left( \oc-i\kc - {\textstyle\frac{1}{2}}\partial_{xx} \right) \photon + \gamma\exciton\,.\label{eq:polaritonB}
\end{align}
\end{subequations}
The coupling constant is half the Rabi frequency $\gamma=\Omega_R/2$; $\omega_X$ is the frequency of a free exciton, $\omega_C$ is the photon frequency at zero wavenumber; and $\kappa_X$ and $\kappa_C$ are the exciton and photon attenuation rates. All these are normalized to a reference frequency $\gamma_0$.
One could set $\gamma_0 = \gamma$, however, we prefer to keep $\gamma$ as an explicit parameter. 
The spatial variable $x$ is normalized to $\ell_0=\sqrt{\hbar/(\gamma_0m_C)\,}$, where $m_C$ is the effective photon mass. 
The wavefunctions $\excitonenv, \photonenv$ are normalised to $\sqrt{N_0}/\ell_0$, 
where $N_0$ is a reference number of particles.
The nonlinearity parameter $g$ is normalised to $N_0/(\ell_0^2\gamma_0)$.
We consider only the case $g>0$ in this paper.
Eqs.~\eqref{eq:polariton} are valid outside the pump spots, at regions that have been the focus
of interesting experimental observations \cite{AmoPigeon_Science2011,TosiChristmannSavvidis_NatPhys2012}.

We seek stationary harmonic polariton fields
\begin{equation}\label{eq:solitonform}
\begin{split}
  \exciton(x,t) = \phi_X(x) e^{-i\omega t} \\
  \photon(x,t) = \photonenv(x) e^{-i\omega t}
\end{split}
\end{equation}
for the lossless equations ($\kappa_X=\kappa_C=0$), with operating frequency $\omega$.
This assumption is reasonable since experimental results \cite{AmoPigeon_Science2011} show 
that the rate of attenuation is slow enough to allow for the formation of solitons.
Letting 
\begin{equation}  \label{eq:wxwc}
  \wx=\omega-\ox,\qquad 
  \wc=\omega-\oc,
\end{equation}
and inserting (\ref{eq:solitonform}) into (\ref{eq:polariton}) yields
\begin{subequations}\label{eq:solitonenvelope}
\begin{align}
  - & {\textstyle\frac{1}{2}}\photonenv'' - \wc\photonenv + \gamma\excitonenv = 0\,, \label{eq:solitonenvelopeA}\\
  & \photonenv = {\textstyle\frac{1}{\gamma}}\left( \wx - g\excitonenv^2 \right)\excitonenv\,. \label{eq:solitonenvelopeB}
\end{align}
\end{subequations}

\begin{figure}[t]
\scalebox{0.28}{\includegraphics{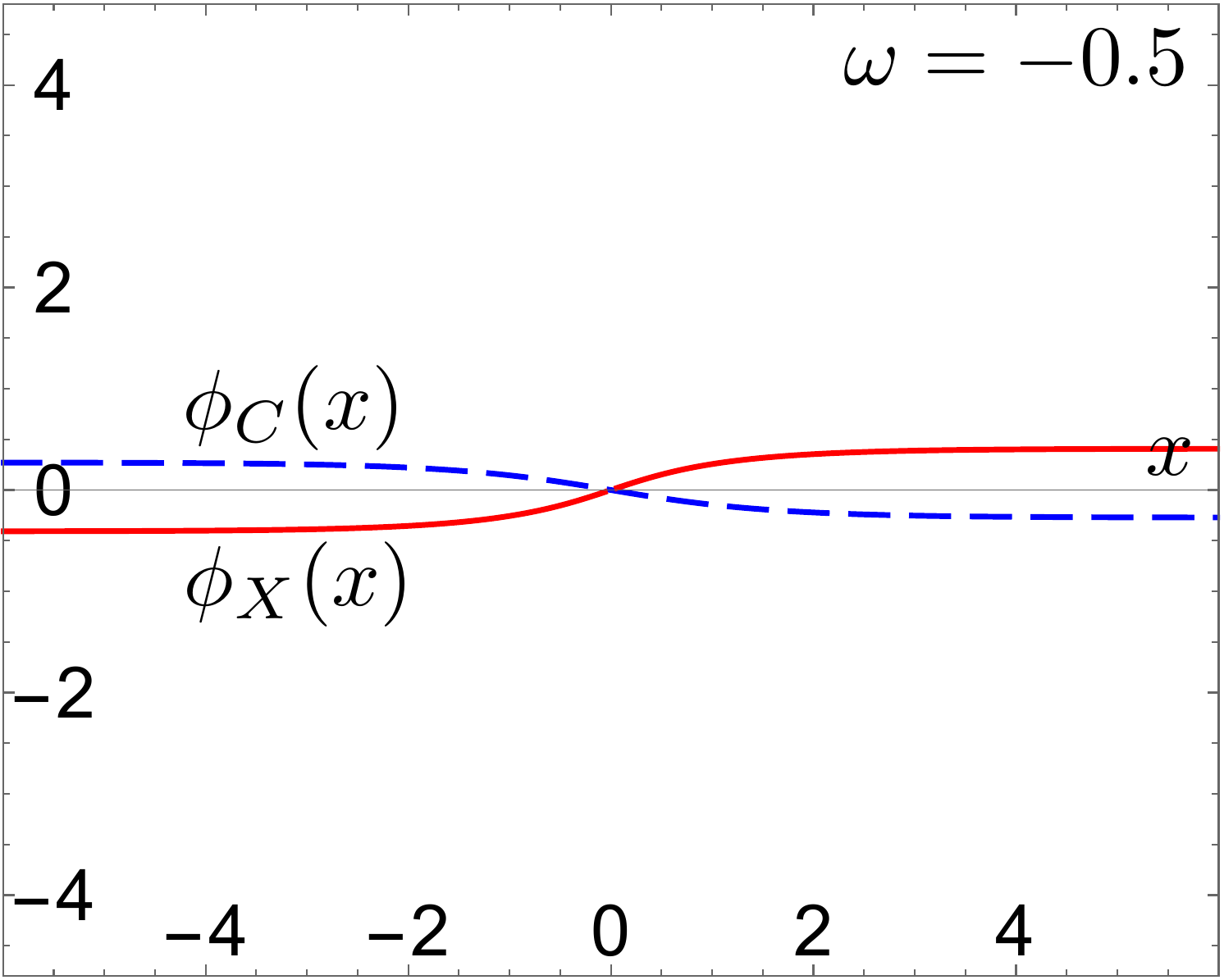}}
\hfill
\scalebox{0.28}{\includegraphics{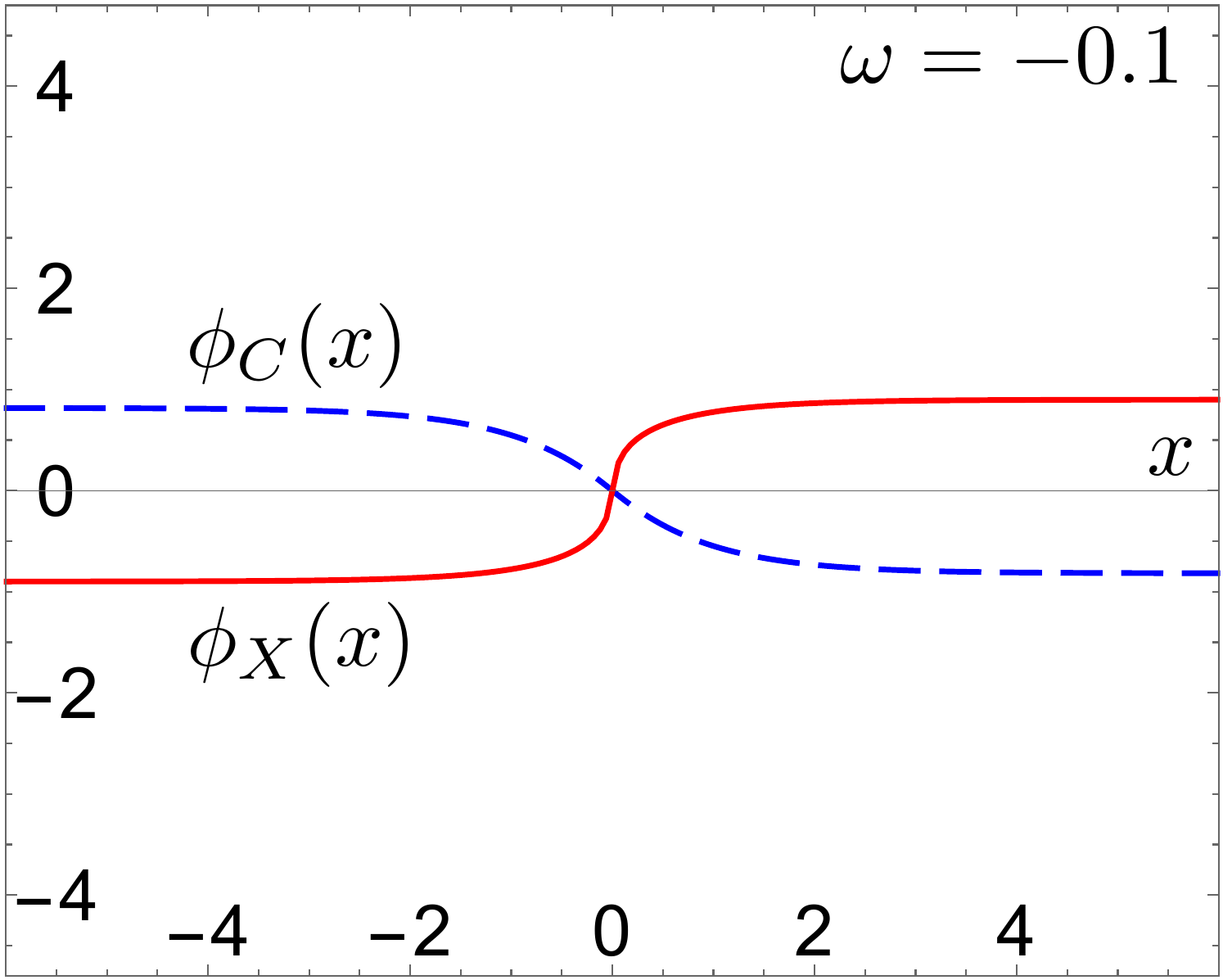}}\\
\scalebox{0.28}{\includegraphics{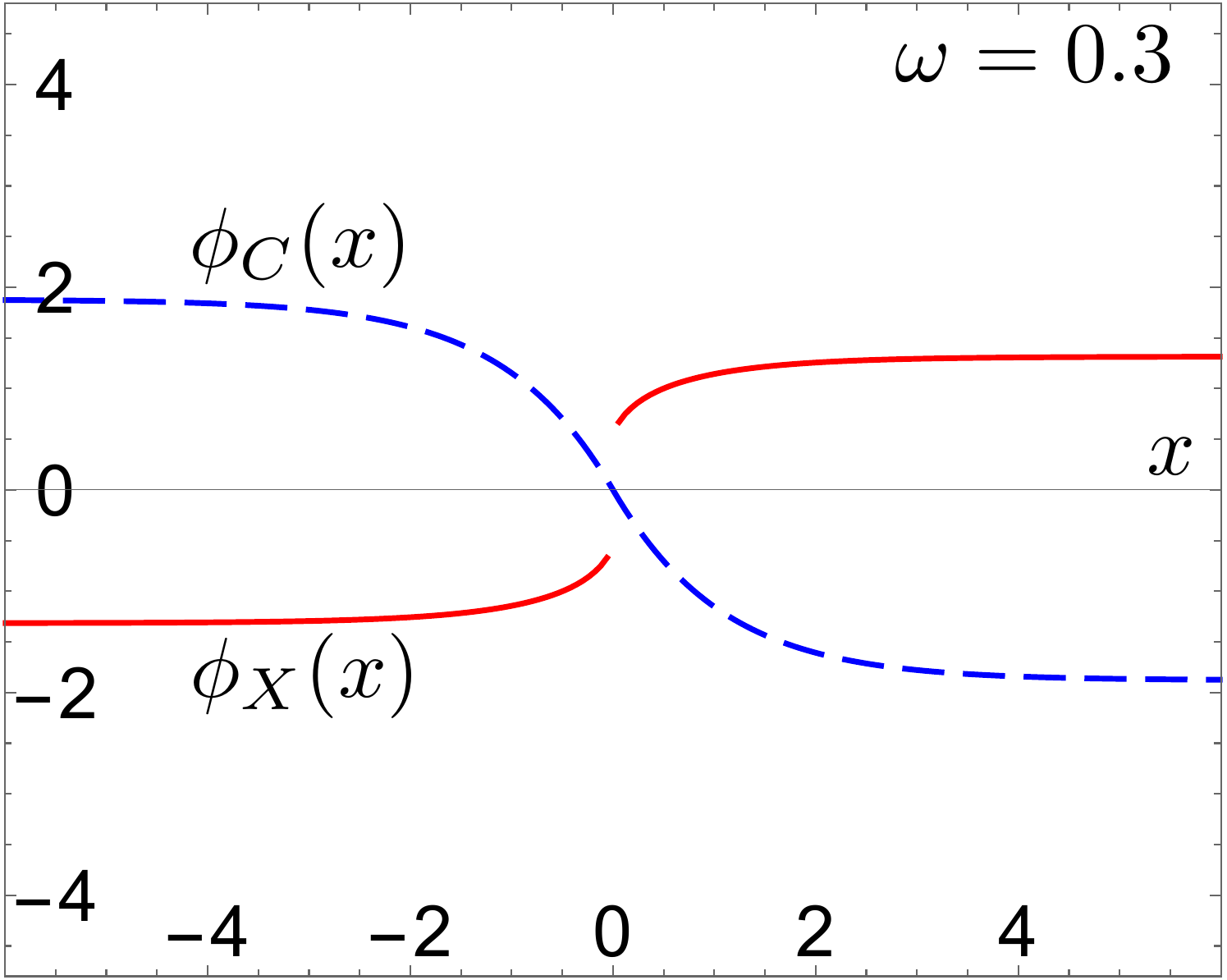}}
\hfill
\scalebox{0.28}{\includegraphics{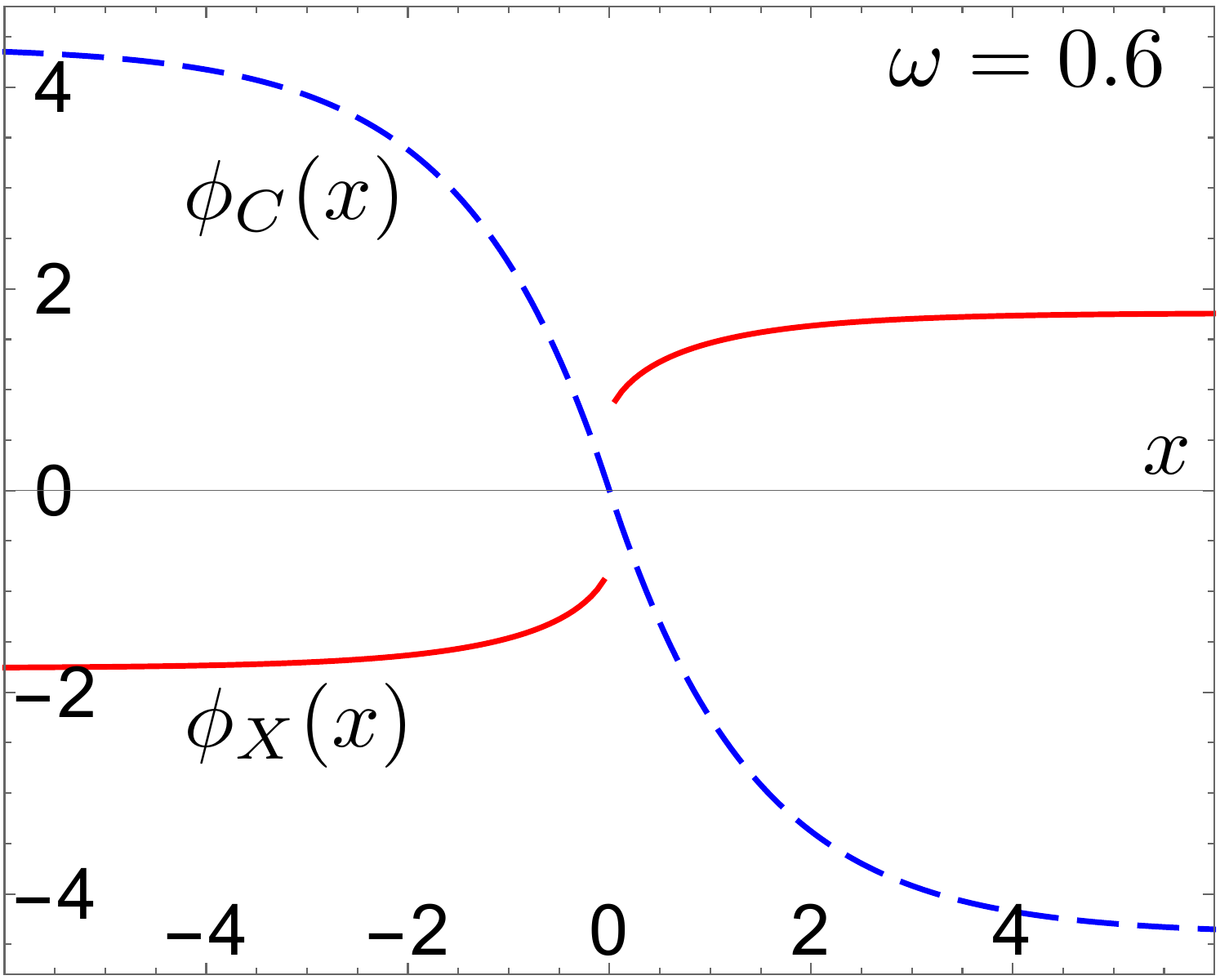}}
   \caption{Dark polariton soliton envelopes $(\excitonenv(x),\photonenv(x))$ for exciton frequency $\ox=0$ and photon frequency $\oc=1$, which gives a threshold frequency $\oLP\approx-0.618$ for the onset of the soliton, a transition frequency $\ox=0$ at which $\excitonenv$ becomes discontinuous, and a blowup frequency $\oc=1$ at which the far-field values of $\excitonenv$ and $\photonenv$ become unbounded as shown in Fig.~\ref{fig:bands} (right).
These graphs demonstrate the increasing soliton amplitude as $\omega$ increases through four values.  When $\omega<\omega_X$, $\excitonenv$ is continuous, and when $\omega>\omega_X$, $\excitonenv$ is discontinuous.
The values of $\excitonenv$ and $\photonenv$ are related by~(\ref{eq:solitonenvelopeB}).
}
\label{fig:solitons}
\end{figure}

Multiplying Eq.~\eqref{eq:solitonenvelopeA} by $\phi_C'$ and  Eq.~\eqref{eq:solitonenvelopeB} by $\gamma\excitonenv'$ and adding the two integrates the system (\ref{eq:solitonenvelope}) exactly.  The cubic algebraic relation (\ref{eq:solitonenvelopeB}) allows one to eliminate $\photonenv$ in favor of $\excitonenv$ to obtain a first-order ODE for $\excitonenv(x)$. It is then convenient to use the scaled exciton density
\begin{equation}
  \zeta(x) := g\,\excitonenv(x)^2
\end{equation}
which eliminates $g$ from the equation and results in
\begin{equation}\label{eq:zetaODE}
\textstyle\frac{1}{2} \zeta'^2 = \dfrac{4\,\zeta Q(\zeta)\,}{(3\zeta-\wx)^2},
\end{equation}
where $Q(\zeta)=-\wc\left[ \zeta^3 - \frac{1}{2} (3\zeta_\infty+\wx)\zeta^2 + \zeta_\infty\wx\zeta + K \right]$, $K$ is an arbitrary real constant of integration, and
\begin{equation}\label{zetainfinity}
  \zeta_\infty = \wx- \frac{\gamma^2}{\wc}
\end{equation}
corresponds to the nonzero equilibrium solution of (\ref{eq:solitonenvelope}).
Eq. \eqref{eq:zetaODE} has the structure of an energy equation of a conservative system and  admits a rich set of solitons and  periodic structures. In this work, we focus on continuous and discontinuous dark solitons for $g>0$.

For a dark soliton $\zeta(x)$ to exist, the cubic polynomial $Q(\zeta)$ must have a double root
that serves as the soliton's far-field value.
The value of the constant of integration $K$ that provides such a nonzero double root equals
\begin{equation}
 K=-\frac{\gamma^6}{2\wc^3} (\eta-1)^2,
\end{equation}
where $\eta$ is a convenient dimensionless parameter
\begin{equation}
 \eta=\frac{(\omega-\omega_X)(\omega-\omega_C)}{\gamma^2}
=\frac{\wx\wc}{\gamma^2}.
\end{equation}
We calculate the double root to be equal to $\zeta_\infty$, given in~\eqref{zetainfinity}.
The fact that this is also the value of the far-field justifies the notation.
As $x$ is varied, $\zeta(x)$ varies continuously down to its minimal value (nadir) $\zeta=0$, which is a simple root  of the potential in \eqref{eq:zetaODE}. 
We may assume that the nadir occurs  at $x=0$.

\begin{figure}[t]
\centerline{
\scalebox{0.33}{\includegraphics{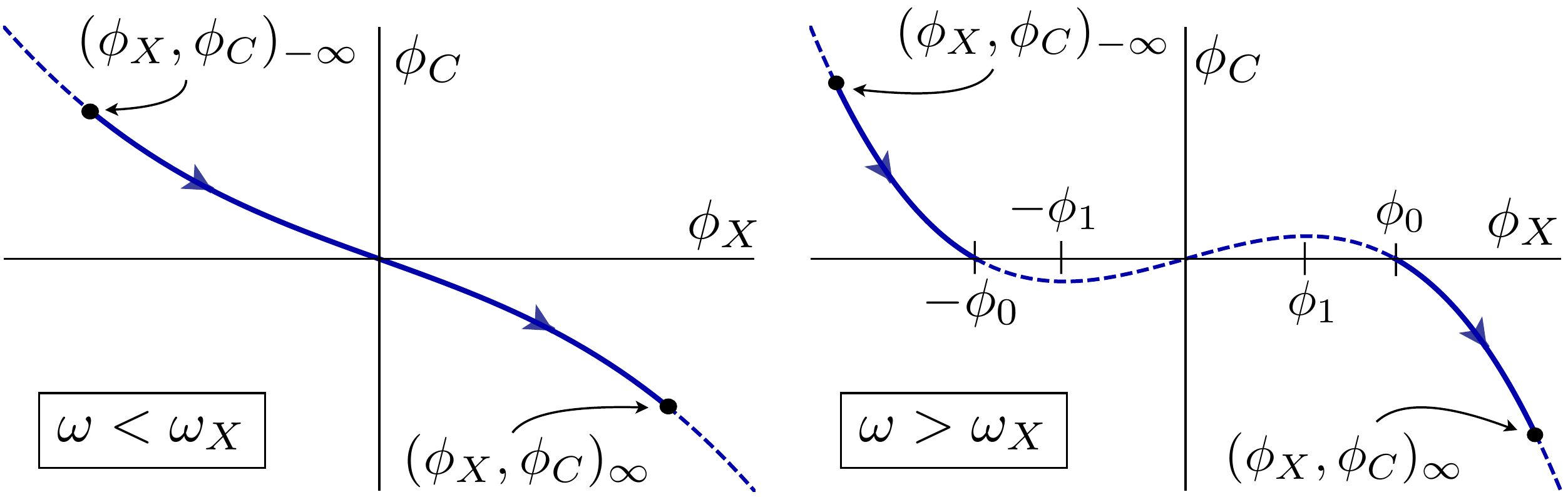}}
}
\caption{The cubic relation (\ref{eq:solitonenvelopeB}) giving the photon field envelope value $\photonenv$ {\itshape vs.} the exciton field envelope value~$\excitonenv$.  {\bfseries Left.} The pair $(\excitonenv(x),\photonenv(x))$ travels continuously along the graph of the monotonic cubic between its far-field values as $x$ increases from $-\infty$ to $\infty$.  {\bfseries Right.} The pair jumps discontinuously between the points $(-\phi_0,0)$ and $(+\phi_0,0)$, with $\phi_0=\sqrt{\wx/g\,}$.  The transition from continuous to discontinuous $\excitonenv$ occurs when $\omega=\omega_X$.  Graphs of the fields $\photonenv(x)$ and $\excitonenv(x)$ are shown in Fig.~\ref{fig:solitons}.
The singularity of the ODE (\ref{eq:zetaODE}) occurs at the critical points $\pm\phi_1$.
}
\label{fig:cubic}
\end{figure}

The soliton field $(\excitonenv(x),\photonenv(x))$ traces the graph of the cubic relation
(\ref{eq:solitonenvelopeB}) as $x$ increases.  Fig.~\ref{fig:cubic} shows the graph of this relation for the two cases $\omega < \ox$ and $\omega > \ox$.  The equilibrium points $({\excitonenv},{\photonenv})_{-\infty}$ and $({\excitonenv},{\photonenv})_{\infty}$
correspond to the calculated value $\zeta_\infty$. 

The parameter $\eta$ is convenient for expressing the soliton {\it nonlinear dispersion relation} at zero wavenumber, that relates the soliton amplitude $\zeta_\infty$
to the operating frequency $\omega$, which is encapsulated in $\eta$ and $\wc$,
\begin{equation}\label{farfield}
 \zeta_\infty= \wx \frac{\eta-1}{\eta} = \frac{\gamma^2}{\wc} (\eta-1). 
\end{equation}
We restrict our attention to $\wc < 0$, which also implies $\eta < 1$, given the fact that $\zeta_\infty > 0$.
Under these conditions, one can show that $Q(\zeta) > 0$, a  necessary condition for Eq.~\eqref{eq:zetaODE}
to have real solutions.

\begin{figure}[t]
\centerline{
\scalebox{0.44}{\includegraphics{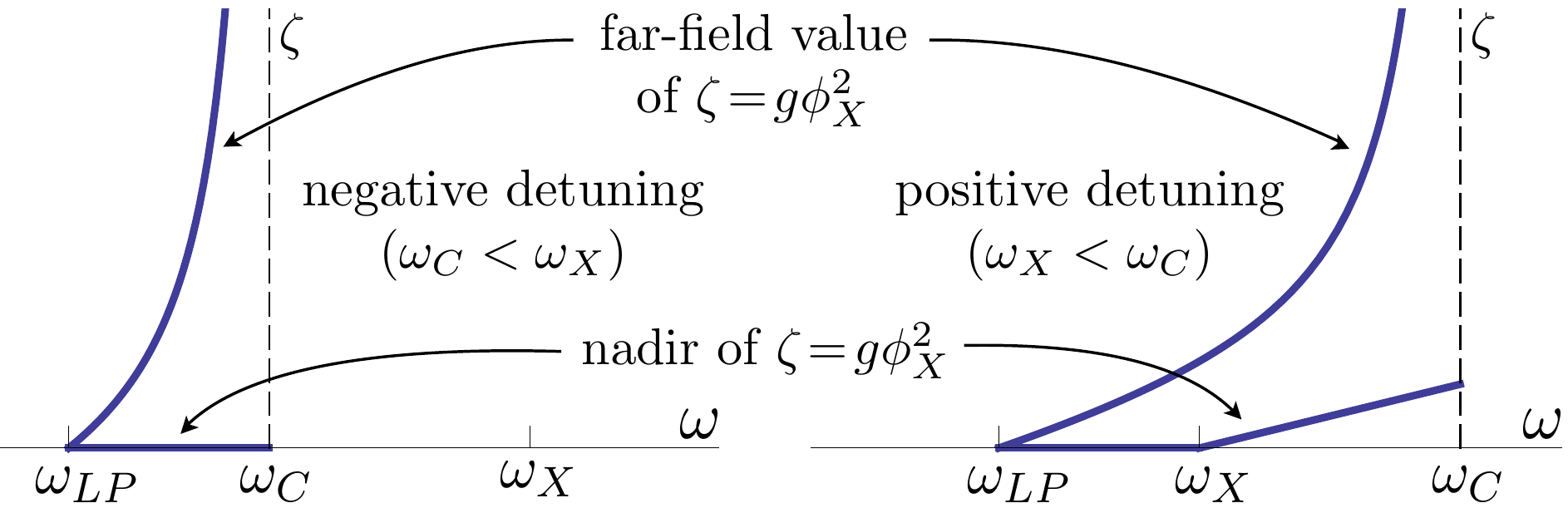}}
}
\caption{ The {\em threshold frequency} $\oLP$ marks the onset of a dark polariton soliton, and the photon frequency $\oc$ is the {\em blowup frequency}, at which the far-field amplitude of the soliton becomes unbounded. {\bf Left}. ($\omega_C<\omega_X$). As the operating frequency $\omega$ traverses the soliton band $(\oLP,\oc)$, the far-field amplitude of the exciton field $\excitonenv$ goes from $0$ to $\infty$ according to \eqref{zetainfinity}.  The nadir (low point) is zero. {\bf Right}. ($\omega_X<\omega_C$) The free exciton frequency $\ox$ is the {\em transition frequency} from continuous to discontinuous solitons.  The nadir of the discontinuous soliton is pushed upwards to the value~$\wx$.}
\label{fig:bands}
\end{figure}

A dark  soliton appears at $\eta=1$ ($\zeta_\infty=0$) corresponding to a {\it threshold frequency}  $\oLP$. This constitutes the linear limit of the  soliton that emerges as the frequency increases; it is thus no surprise that the frequency $\oLP$  coincides with the lower endpoint of the well-known lower band $(\oLP,\ox)$ of homogeneous linear $(g=0)$ polaritons of the form $(\excitonenv,\photonenv)e^{i(kx-\omega t)}$, with $\excitonenv$ and $\photonenv$ constant~\cite{MarchettiSzymanska}.
As the frequency is increased from its threshold, the value of $\eta$ decreases and the amplitude of the soliton increases until it blows up at the photon frequency $\oc$ ($\eta=0$, \ $\zeta_\infty=\infty$). Fig.~\ref{fig:bands} displays the far-field and nadir values of the soliton {\it vs.}\;the frequency in the band from threshold to blowup, in the cases of negative detuning and positive detuning.

In the case of positive detuning, $\omega_X\!<\!\omega_C$, ({\it i.e.} $\wc<\wx$), the frequency 
$\omega_X$ lies within the soliton frequency band $(\oLP,\oc)$, constituting a {\em transition frequency}  above which the soliton field $\excitonenv$ becomes discontinuous.    The obstructing singularity $\zeta=\wx/3$ becomes positive, breaking  into the soliton range $[0,\zeta_\infty)$. The nadir of the soliton is pushed upward  from $\zeta=0$ to the value $\zeta=\wx$, which is now positive,  leading to a jump of the exciton field  between the values $\pm\phi_0=\pm\sqrt{\wx/g}$.  Fig.~\ref{fig:cubic} traces the path of the pair $(\excitonenv(x),\photonenv(x))$ along the graph of the relation (\ref{eq:solitonenvelopeB}) both for negative detuning and positive detuning. The system equations \eqref{eq:solitonenvelope} remain valid, as the jump in $\excitonenv$ is balanced by a jump in~$\photonenv''$. Physically, the photon field $\photonenv$ which mediates the coupling between neighboring excitons through the term $\gamma\photonenv$ in (\ref{eq:polaritonA}), vanishes when $\zeta$ takes the special value $\wx$ (corresponding to $\phi_0=\sqrt{\wx/g}$ in Fig.~\ref{fig:cubic}b). The vanishing of the photon field  turns off the coupling between neighboring excitons  thus making the jump permissible.   The formulae \eqref{zetainfinity} and \eqref{farfield}
for the far-field value $\zeta_\infty$ remain the same.

Fig.~\ref{fig:solitons} presents four instances of the soliton profile that show the progress towards
the discontinuity (top) and the progress past the discontinuity of the exciton field (bottom).
The photon field remains continuous.
Its second derivative has a discontinuity at $x=0$, as discussed earlier, but this is
not visible in the figure.  Notice the monotonic increase of the far-field amplitude as the frequency $\omega$ increases.

\section{Healing Length and Comparison with Gross-Pitaevskii equation}

It is interesting to visualize the mechanism of the formation of the discontinuity of the exciton field $\excitonenv(x)$ by following the slope of this field 
at  $x=0$, as one lowers the dimensionless parameter  $\eta$ from its value $\eta=1$ at which the dark soliton is born. In order to calculate this slope, we  express $|\phi'_X(0)|$ in terms of  $\zeta$ and $\zeta'$ from the relation
$\zeta=g\excitonenv^2$. We then   insert the value  for  $\zeta'$ from  the differential equation \eqref{eq:zetaODE}  and, finally,  set  $\zeta=0$. 
We obtain  
\begin{equation}\label{Xslope}
 \left[ \phi'_X(0) \right]^2 = \frac{\gamma^2 (\eta-1)^2 }{g\,\eta^2}.
\end{equation}
For positive detuning  and as $\omega\nearrow \omega_X$, \ the parameter  $\eta\searrow0$ and thus, the slope $\excitonenv'(0)$ tends to infinity, while $\excitonenv$ remains finite.  The jump discontinuity of the exciton envelope profile sets on as $\eta$ becomes negative.

Adopting the slope of the profile at the origin $x=0$ as an
indicator of the scale of the slope
of the profile we define the {\em healing length} of a exciton field profile by 
\begin{equation}
 \xi_X \,=\, 2\,\left|\frac{\excitonenv(x=\pm\infty)}{\excitonenv'(0)}\right|\,,
\end{equation}
with a similar equation for the photon field.
From the field envelope Eq.~\eqref{eq:solitonenvelopeA}, and the far-field Eq.~\eqref{farfield}, we obtain 
$\photonenv(\infty)/\excitonenv(\infty)=\gamma/\wc$ and $\photonenv'(0)/\excitonenv'(0)=\wx/\gamma$. Thus, the healing lengths $\xi_C$ and  $\xi_X$  are related~by  
\begin{equation}
 \xi_C^2=\frac{\xi_X^2}{\eta^2}.
\end{equation}
Combining Eqs.~\eqref{farfield},  \eqref{Xslope} and $\zeta_\infty=g\phi_\infty^2$,   we obtain for the continuous soliton the healing lengths 
\begin{equation}\label{eq:healing_length_Full_system}
 \xi_X^2=\frac{4\eta^2}{\wc (\eta-1)}, \ \ \ 
\xi_C^2=\frac{4}{\wc (\eta-1)}. 
\end{equation}

When $\oc<\ox$, near the blow-up frequency $\wc=0\, (\eta=0)$ the healing length of the excitons approaches zero,
while the photon healing length diverges to infinity. At the same time the far-field value goes to infinity.
At the transition frequency $\wx=0\, (\eta=0)$ (obtained only for positive detuning) $\xi_X$ goes to
zero linearly in $\eta$ which one can view as a precursor to the discontinuity. 
The photon healing length converges to $\xi_C^2=4/(\oc-\ox)$.
Fig.~\ref{fig:solitons} exemplifies these observations.

In the region near the value $\eta=1$, at which the  continuous soliton begins its life, the exciton and the photon fields are nearly proportional to each other and $\xi_C \approx \xi_X$.
The photon field is described well by a Gross-Pitaevskii (GP) model that is  derived as a simplification of the two-equation model \eqref{eq:solitonenvelope}.
 We solve Eq. \eqref{eq:solitonenvelopeB} for $\phi_X$ as a power series in  $\photonenv$ up to the third degree term and we insert this value of  $\phi_X$ into Eq. \eqref{eq:solitonenvelopeA}.
There seems to be no analogous way to derive a GP equation for the exciton field.
The GP model derived for the photon field is
 \begin{equation}  \label{eq:GP}
\textstyle\frac{1}{2} \phi''-\varepsilon \wc\,\phi-\tilde g\phi^3=0.
\end{equation}
The notation $\phi$ is a convenient abbreviation of the more descriptive notation $\phi_{\rm GP,C}$.
The parameter $\varepsilon>0$ measures the deviation from the linear problem
and equals
\begin{equation}
 \varepsilon=\dfrac{1-\eta}{\eta},
\end{equation}
while $\tilde{g} = (\textstyle\frac{\wc}{\wx})^2g$.  
 
Multiply by $2\phi'$ and integrate to obtain 
\begin{equation}\label{eq:GP_conservation_law}
 \textstyle\frac{1}{2}\phi'^2-\varepsilon \wc\, \phi^2- \textstyle\frac{1}{2}\tilde g\phi^4=E_0
\end{equation}
where $E_0$ is a constant of integration.  Like Eq.~\eqref{eq:zetaODE}, this has the structure
of a conservative system. The left side can be considered as the sum of a kinetic and a potential energy.
It produces the GP approximation of the photon profile of the soliton we are investigating.
The potential has two equal maxima at
$\pm\phi_\infty$ where 
\begin{equation}
 \phi_\infty^2=-\frac{\varepsilon \wc}{\tilde g}.
\end{equation}
These are the far-field values ($\phi'=0$) for soliton solutions obtained from Eq.~\eqref{eq:GP_conservation_law}
at the peak of the potential
\begin{equation}
 E_0= -\varepsilon \wc \phi_\infty^2- \textstyle\frac{1}{2}\tilde{g}\phi_\infty^4=-\dfrac{1}{2} \varepsilon \wc \phi_\infty^2. 
\end{equation}

We obtain from Eq. \eqref{eq:GP_conservation_law}
$\phi'(0)^2=2E_0=-\varepsilon \wc \phi_\infty^2.$
Taking, as before,  the slope  $|\phi'(0)|$ as an indicator of the slope of the profile,
the healing length for the photons is 
\begin{equation}
\label{eq:healing_length_GP}
 ( \xi_C^{\mbox{\scriptsize GP}})^2=\frac{4\phi(\pm\infty)^2}{\phi'(0)^2}=\frac{4\phi_\infty^2}{\phi'(0)^2}
 = \dfrac{4\eta}{\wc (\eta-1)}.
\end{equation}
The photon healing length for the approximate equation (GP) underestimates the healing length
derived for the full system in \eqref{eq:healing_length_Full_system} by a factor of $\eta$. 
The two agree at the linear limit $\eta=1$.

\section{Soliton as a photon field in a potential well}

Returning to the system involving both the photon and the exciton fields,
one can write Eqs.~(\ref{eq:solitonenvelope}) as a Schr\"odinger equation for the photon field envelope $\photonenv$,
\begin{equation}\label{SchrodingerPhoton}
  -{\textstyle\frac{1}{2}}\photonenv'' + V(x)\photonenv \,=\, \wc\photonenv\,,
\end{equation}
in which the effective potential $V(x)$ depends on the exciton field:
\begin{equation}\label{potential}
  V(x) = \frac{\gamma^2}{\wx-g\excitonenv(x)^2}\,.
\end{equation}

\begin{figure}
\scalebox{0.28}{\includegraphics{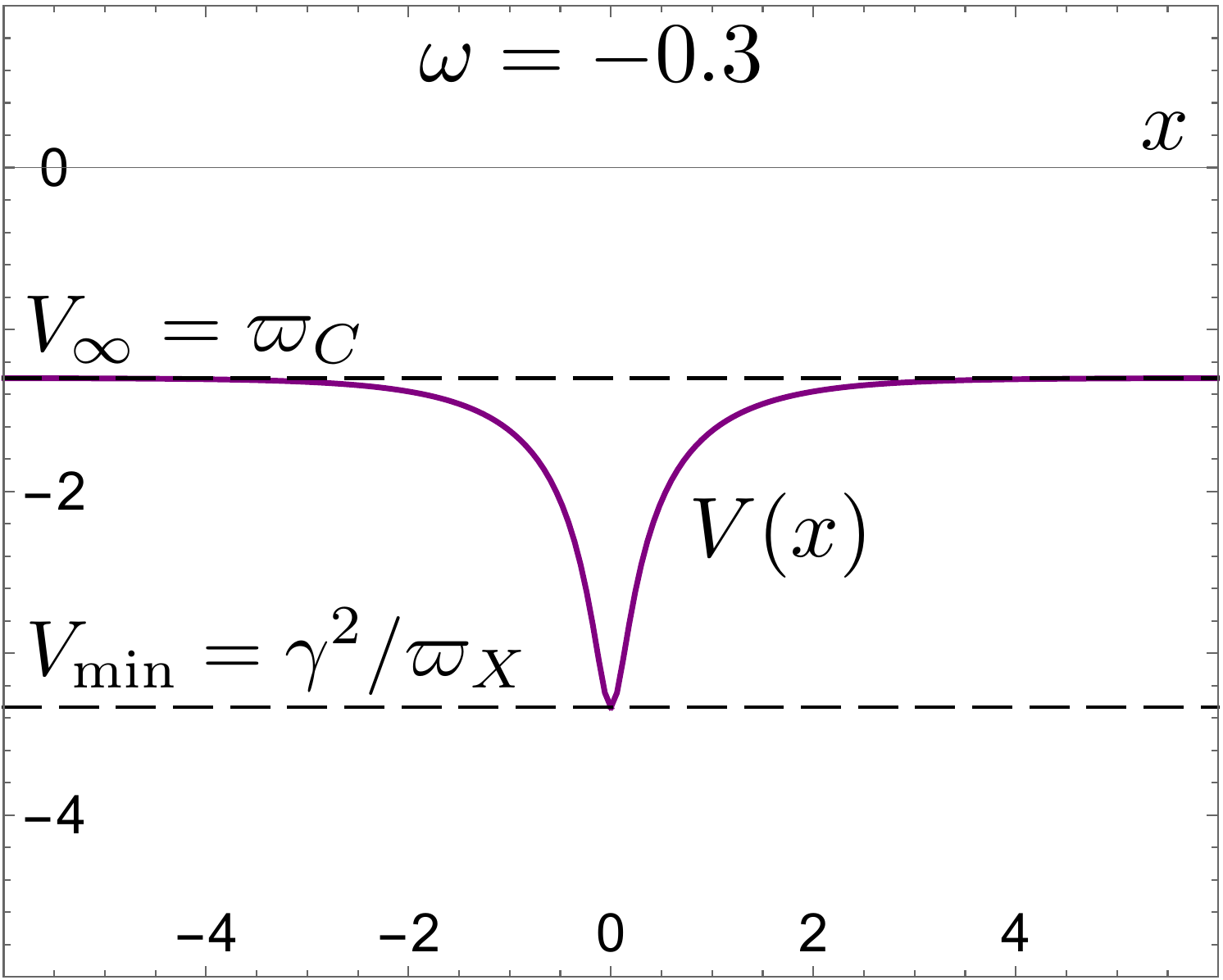}}
\scalebox{0.28}{\includegraphics{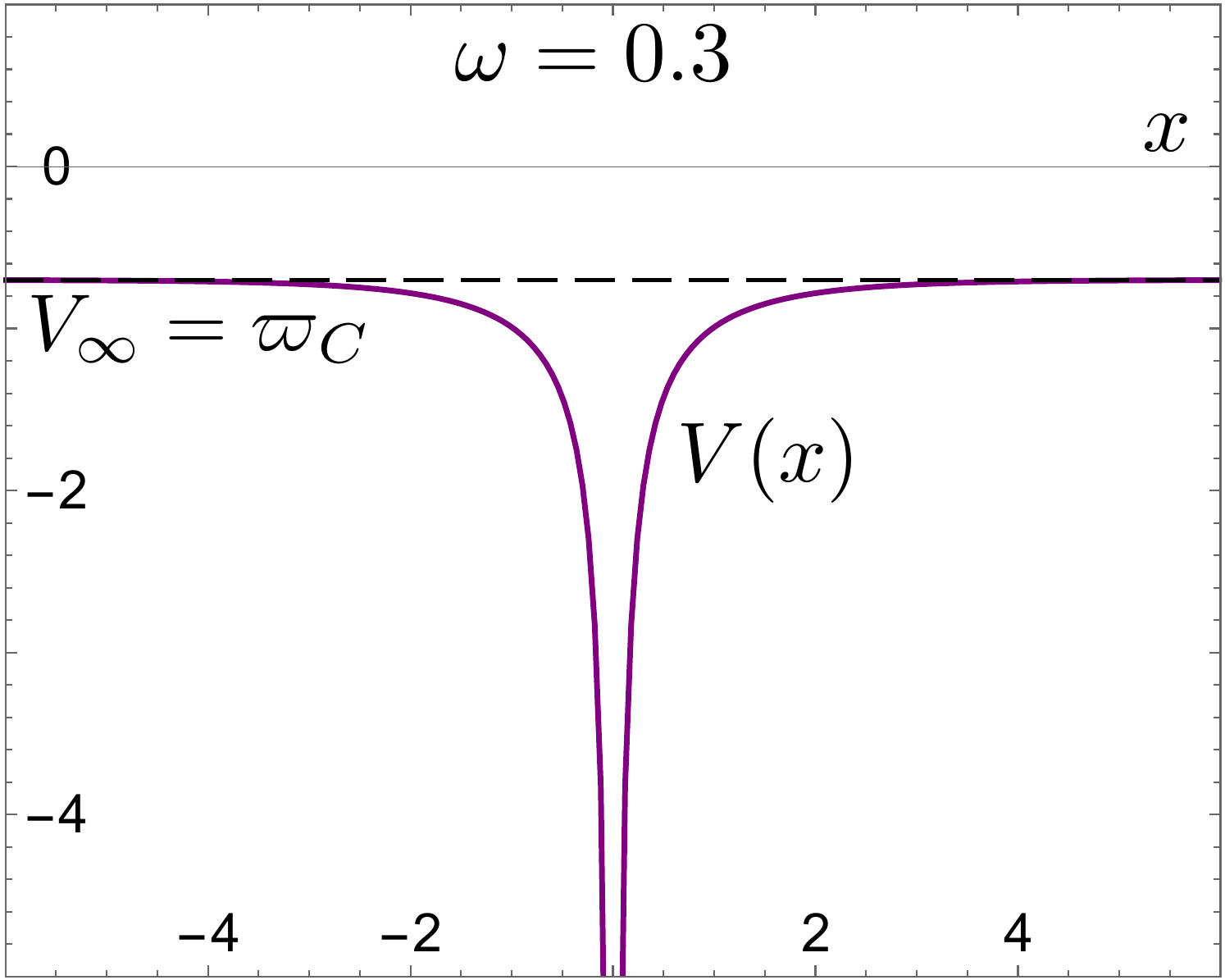}}
\caption{The effective potential well $V(x)$ that confines the photon field of an exciton-polariton soliton.  {\bfseries Left.} When the exciton field is continuous ($\omega<\omega_X$), $V(x)$ has a finite minimal value.  {\bfseries Right.} When the exciton field is discontinuous ($\omega>\omega_X$), $V(x)$ is unbounded at the point of discontinuity $x=0$. ($\ox=0$, $\oc=1$, and $\oLP\approx-0.618$, as in Fig.~\ref{fig:solitons}.)}
\label{fig:potential}
\end{figure}

For the dark soliton derived above, $V(x)$ exhibits a single symmetric well with far-field value $V_\infty = \wc<0$, as shown in Fig.~\ref{fig:potential}.  For the continuous soliton, $V$ has a minimal value of $V_\mathrm{min} =  \gamma^2/\wx$.  For the discontinuous soliton, the well becomes infinitely deep at the point of discontinuity.
%

In an experimental setup, one expects that losses will allow some photons to be trapped by 
the potential well \eqref{potential} in the form of bound states at discrete energy levels which lie below $\wc$.
 As long as a small enough fraction of the energy of the photon field of the coherent polariton structure is transferred into lower energy states, the exciton field $\excitonenv(x)$ and therefore also the potential $V(x)$ will not be significantly altered and can be considered a fixed potential.

This scenario is consistent with experimental observations \cite{TosiChristmannSavvidis_NatPhys2012}, in which a polariton field is sustained by continuously injecting photons at two pump spots, one on each side of the potential well.  
A fraction of the polariton population descends to lower energy states of the well.

\section{Conclusions}

We have presented a detailed study of dark solitons in polariton condensates,
which result as solutions of a system of equations for strongly coupled excitons and photons.
We have analytically identified soliton solutions for the lossless system.
One type of black soliton studied is of the standard type where the fields vanish at the soliton center.
This corresponds to complete depletion of the condensate at that point.
Furthermore, we reported a discontinuous soliton where the exciton field
exhibits a jump at the soliton center, so the exciton density does not vanish.
We have shown that the two types of solitons can be unified in one brach,
since the discontinuity in the exciton density smoothly increases from zero.

Polariton condensates emerge as a fertile ground for solitonic structures.
Our results provide an understanding of these structures.
Furthermore, they can be used as a basis for a perturbation theory that will include non-conservative
features, in particular, sources and losses.

\acknowledgements
This work was partially supported by the European Union's FP7-REGPOT-2009-1 project 
``Archimedes Center for Modeling, Analysis and Computation'' (grant agreement n. 245749),
by the (US) National Science  Foundation  under grants NSF DMS-0707488 and NSF DMS-1211638,
and by EU and Greek national funds through the Operational Program ``Education and Lifelong Learning'' -- THALES.
This work has benefited from discussions with P. Savvidis, G. Christmann, F. Marchetti, G. Kavoulakis, A. Gorbach.

\end{document}